\documentclass[12pt]{article}
\usepackage{a4wide}
\usepackage{latexsym}
\usepackage{amsmath}
\usepackage{amsfonts}
\usepackage{amsthm}
\usepackage{amscd}
\usepackage{cite}
\usepackage{graphicx}

\usepackage{pslatex}
\usepackage[latin1,utf8]{inputenc}
\usepackage[T1]{fontenc}

\usepackage{ifthen}

\allowdisplaybreaks

\newcommand{\bq}{\begin{eqnarray}}
\newcommand{\eq}{\end{eqnarray}}
\newcommand{\eps}{\varepsilon}

\begin{document}

\thispagestyle{empty}

\begin{flushright}
  MITP/18-011
% \\ version of \today
\end{flushright}

\vspace{1.5cm}

\begin{center}
  {\Large\bf The $\varepsilon$-form of the differential equations for Feynman integrals in the elliptic case \\ 
  }
  \vspace{1cm}
  {\large Luise Adams and Stefan Weinzierl \\
  \vspace{1cm}
      {\small \em PRISMA Cluster of Excellence, Institut f{\"u}r Physik, }\\
      {\small \em Johannes Gutenberg-Universit{\"a}t Mainz,}\\
      {\small \em D - 55099 Mainz, Germany}\\
  } 
\end{center}

\vspace{2cm}

% abstract ---------------------------------------
\begin{abstract}\noindent
  {
Feynman integrals are easily solved if their system of differential equations is in $\varepsilon$-form.
In this letter we show 
by the explicit example of the kite integral family 
that an $\varepsilon$-form can even be achieved, if the Feynman integrals do not evaluate
to multiple polylogarithms.
The $\varepsilon$-form is obtained by a (non-algebraic) change of basis for the master integrals.
   }
\end{abstract}

\vspace*{\fill}

% -----------------------------------------------------------------------------
\newpage

\section{Introduction}
\label{sec:intro}

Precision calculations in high-energy physics rely on our ability to compute Feynman loop integrals.
In recent years, the method of 
differential equations \cite{Kotikov:1990kg,Kotikov:1991pm,Remiddi:1997ny,Gehrmann:1999as,Argeri:2007up,MullerStach:2012mp,Henn:2013pwa,Henn:2014qga,Ablinger:2015tua,Bosma:2017hrk}
has become a powerful computational tool for these integrals.
This is in particular true, if the system of differential equations is in $\eps$-form (where $\eps$ denotes the regularisation parameter
of dimensional regularisation) \cite{Henn:2013pwa}.
The Feynman integrals under consideration are then easily expressed in terms of iterated integrals
of the integration kernels appearing on the right-hand side of the differential equations.
In the past this method has been applied very successfully to Feynman integrals which evaluate to multiple polylogarithms.
In this case, the integration kernels, when written as one-forms are given by
\bq
 \frac{dt}{t-z_i},
\eq
where the $z_i$'s are (possibly complex) parameters, called the letters of the multiple polylogarithms.
Thus, the computation of Feynman integrals is reduced to finding a transformation, which brings the system
of differential equations to $\eps$-form \cite{Gehrmann:2014bfa,Argeri:2014qva,Lee:2014ioa,Prausa:2017ltv,Gituliar:2017vzm,Meyer:2016slj,Adams:2017tga,Lee:2017oca,Meyer:2017joq,Becchetti:2017abb}.
In the case where the Feynman integrals evaluate to multiple polylogarithms, one considers rational or algebraic transformations.

In this letter we would like to point out, that the fact that a system of differential equations for Feynman integrals
can be brought to an $\eps$-form is not restricted to Feynman integrals which evaluate to multiple polylogarithms.
We do this by giving an explicit example.
We consider the Feynman integrals associated to the kite integral.
In this family of Feynman integrals we have an elliptic sub-sector, given by the sunrise integral.
The $\eps$-form is obtained by a suitable change of basis for the master integrals.
In contrast to the multiple polylogarithm case, this transformation is not rational or algebraic, but transcendental.
For example, the new first master integral in the sunrise sector is given as the original sunrise integral divided by a period
of the elliptic curve, the latter being a transcendental function of the kinematic invariants.
It is the extension of basis transformations from algebraic/rational transformations to
transcendental transformations which allows us to obtain an $\eps$-form.
On the other hand, the change of basis of master integrals is natural:
Let us stay with the example of the first master integral in the sunrise sector:
The new first master integral in this sector is nothing than the original integral divided by the $\eps^0$-term of the maximal cut for a specific contour,
a procedure also often used in the multiple polylogarithm case to obtain master integrals of uniform weight.
Part of the results presented here follow from our earlier work 
on the sunrise/kite-integrals \cite{Broadhurst:1993mw,Berends:1993ee,Bauberger:1994nk,Bauberger:1994by,Bauberger:1994hx,Caffo:1998du,Laporta:2004rb,Kniehl:2005bc,Groote:2005ay,Groote:2012pa,Bailey:2008ib,MullerStach:2011ru,Adams:2013nia,Bloch:2013tra,Adams:2014vja,Adams:2015gva,Adams:2015ydq,Remiddi:2013joa,Bloch:2016izu,Sabry:1962,Remiddi:2016gno,Adams:2016xah,Adams:2017ejb,Bogner:2017vim},
and in particular from refs.~\cite{Adams:2016xah,Adams:2017ejb}.
The transition to the $\eps$-form is made possible by a suitable definition of the second master integral in the sunrise sector.
On a technical level, this is the essential new result of this article.

Apart from finding an $\eps$-form for the system of differential equations we may also ask if the integration kernels
have ``nice'' properties.
In the case of multiple polylogarithms they are simple rational functions, related to ``$\mathrm{dlog}$''-forms
\bq
 \frac{dt}{t-z_i}
 & = &
 d \ln\left(t-z_i\right).
\eq
For the kite integral and the sunrise integral we find that after a change of variables from $x=p^2/m^2$ to the modular
parameter $\tau$, all our integration kernels are modular forms.
Thus, for the example of the integral family of the kite integral, all Feynman integrals belonging to this family
may be expressed as iterated integrals of modular forms.
The required change of variables is not algebraic/rational but transcendental.
But again, it is quite natural: The new variable $\tau$ parametrises the moduli space of the family of elliptic
curves.

This paper is organised as follows:
In section~\ref{sec:diff_eq} we review the method of differential equations. 
We may view a system of differential equations of Feynman integrals
as a fibre bundle with a Gau{\ss}-Manin connection.
We discuss the effects of a change of the basis of the fibre and the effects of a change of coordinates of the
base manifold.
In section~\ref{sec:eps_form} we give an explicit example 
for the $\eps$-form of a system of differential equations for Feynman integrals in the elliptic case.
Finally, section~\ref{sec:conclusions} contains our conclusions.
The appendix contains useful information on Eisenstein series, dimensional shift relations and the second master integral in the sunrise sector.

% -----------------------------------------------------------------------------
\section{Differential equations}
\label{sec:diff_eq}

We consider Feynman integrals within dimensional regularisation
and we denote the dimensional regularisation parameter by $\eps$.
Let $I_i \in \{I_1,...,I_N\}$ be a master integral and 
let $x_\alpha$ be a dimensionless kinematic variable 
(for example an external Lorentz invariant $x_\alpha=(p_{a_\alpha}+p_{b_\alpha})^2/\mu^2$ 
or an internal mass $x_\alpha=m_\alpha^2/\mu^2$) .
Carrying out the derivative $\partial I_i/\partial x_\alpha$ 
under the integral sign and using integration-by-parts identities allows us to express the 
derivative as a linear combination of the master integrals:
\bq
 \frac{\partial}{\partial x_\alpha} I_i
 & = &
 \sum\limits_{j=1}^N A^\alpha_{ij} \; I_j
\eq
More generally, let us denote by $\vec{I} = \left(I_1,...,I_N\right)$ the vector of the master integrals,
and by $\vec{x} = \left(x_1,...,x_n\right)$ the vector of independent kinematic variables the master integrals depend on.
Repeating the above procedure for every master integral and every kinematic variable we 
obtain a system of differential equations of Fuchsian type
\bq
\label{diff_eq_original}
 d\vec{I} & = & A \; \vec{I},
\eq
where $A$ is a matrix-valued one-form
\bq
 A_{i j} & = & 
 \sum\limits_{\alpha=1}^n A^\alpha_{i j} \; dx_\alpha,
 \;\;\;\;\;\;\;\;\; 
 i,j \in \{1,...,N\},
\eq
satisfying the integrability condition $dA - A \wedge A = 0$.
In general, the matrix $A$ depends on $\vec{x}$ and the dimensional regularisation parameter $\eps$: 
\bq
 A & = & A\left(\vec{x},\eps\right).
\eq
We say that the system of differential equations is in $\eps$-form, if eq.~(\ref{diff_eq_original}) is of the form
\bq
\label{diff_eq_eps_form}
 d\vec{I} & = & \eps \; A' \; \vec{I},
\eq
where the matrix-valued one-form $A'$ is independent of $\eps$:
\bq
 A' & = & A'\left(\vec{x}\right).
\eq
Once the system of differential equations is in $\eps$-form, it is easily solved in terms of iterated integrals
of the one-forms appearing in the entries of the matrix $A'$.
Thus the class of functions required to express the Feynman integrals $\vec{I}$ is exactly given by
the iterated integrals of the one-forms appearing in the entries of the matrix $A'$.

Mathematically, we are considering a vector bundle, where $\vec{x}$ denotes the coordinates of the base manifold,
$\vec{I}$ defines a basis of the fibre and
\bq
 \nabla_A & = & d - A
\eq
is the flat Gau{\ss}-Manin connection.

Let us first discuss a change of variables for the base manifold:
\bq
\label{trafo_base}
 x_\alpha' & = & f_\alpha\left(\vec{x}\right)
\eq
We denote the Jacobian by
\bq
 J_{\alpha \beta} & = & \frac{\partial x_\alpha'}{\partial x_\beta}.
\eq
Then the differential equations
\bq
 \frac{\partial I_i}{\partial x_\alpha}
 & = &
 \sum\limits_{j=1}^N
 A^\alpha_{i j} \; I_j
\eq
transform into
\bq
 \frac{\partial I_i}{\partial x_\alpha'}
 & = &
 \sum\limits_{j=1}^N
 \left(A'\right)^\alpha_{i j} \; I_j,
 \;\;\;\;\;\;\;\;\;
 \left(A'\right)^\alpha_{i j}
 \; = \; 
 \sum\limits_{\beta=1}^n A^\beta_{i j} \left(J^{-1}\right)_{\beta \alpha}
\eq
A change of variables for the base manifold is often performed to eliminate square roots.
For example, the transformation \cite{Fleischer:1998nb,Kotikov:2007vr,Bonciani:2010ms,Henn:2013woa}
\bq
\label{example_base_change}
 x' \;\; = \;\;
 \frac{1}{2} \left( 2-x - \sqrt{-x} \sqrt{4-x} \right),
 & &
 x \;\; = \;\; -\frac{\left(1-x'\right)^2}{x'},
\eq
leads to
\bq
 \frac{dx}{\sqrt{-x}\sqrt{4-x}}
 & = &
 \frac{dx'}{x'}.
\eq
Eq.~(\ref{example_base_change}) is an algebraic transformation.

Secondly, we may also change the basis in the fibre: 
Let us denote by
\bq
\label{trafo_fibre}
 \vec{I}\;{}' & = & U \; \vec{I}
\eq
a new basis of the master integrals.
Then the differential equation becomes
\bq
 d \vec{I}\;{}' 
 & = & 
 A' \; \vec{I}\;{}',
 \;\;\;\;\;\;\;\;\;
 A' = U A U^{-1} - U d U^{-1}.
\eq
In the sequel we will always be considering linear transformations as in eq.~(\ref{trafo_fibre}).
We call such a transformation rational, algebraic or transcendental, if the entries of the transformation matrix $U$
are rational functions of $\vec{x}$, algebraic functions of $\vec{x}$ or transcendental functions of $\vec{x}$, respectively.
Typical algorithms \cite{Lee:2014ioa,Prausa:2017ltv,Gituliar:2017vzm,Meyer:2016slj,Lee:2017oca,Meyer:2017joq}
search for a transformation matrix $U$ where the entries are rational functions of $\vec{x}$.

Thus, up to now the typical strategy has been to combine algebraic/rational transformations of 
the form of eq.~(\ref{trafo_base}) and eq.~(\ref{trafo_fibre}) to transform the system of differential equations
into an $\eps$-form as in eq.~(\ref{diff_eq_eps_form}), where the matrix $A'$ has the form
\bq
\label{dlog_form}
 A' & = & \sum\limits_{r=1}^k \; C_r \; d\ln p_r(\vec{x}),
\eq
with the $C_r$'s being $N \times N$-matrices with constant entries,
the $p_r$'s being polynomials in $\vec{x}$ and $k$ being the number of those polynomials.
If the system of differential equations can be transformed to an $\eps$-form with $A'$ as in eq.~(\ref{dlog_form}),
the Feynman integrals are expressible in terms of multiple polylogarithms.

However, there are Feynman integrals which cannot be expressed in terms of multiple polylogarithms.
In this letter we show with an explicit example that even in this case the system of differential equations 
may be transformed to the $\eps$-form of eq.~(\ref{diff_eq_eps_form}).
It is clear that one cannot achieve the specific ``dlog''-form of eq.~(\ref{dlog_form}), because otherwise 
the Feynman integrals would be expressible in terms of multiple polylogarithms.
The transformation to the $\eps$-form uses only the transformations of
eq.~(\ref{trafo_base}) and eq.~(\ref{trafo_fibre}), however we do not require that these transformations are algebraic or
rational.

% -----------------------------------------------------------------------------
\section{An example of the $\eps$-form of the differential equations for Feynman integrals in the elliptic case}
\label{sec:eps_form}

The integral family for the kite integral is given in $D$-dimensional Minkowski space by 
\bq
\label{def_kite}
 I_{\nu_1 \nu_2 \nu_3 \nu_4 \nu_5}\left( D, p^2, m^2, \mu^2 \right)
 & = &
 \left(-1\right)^{\nu_{12345}}
 \left(\mu^2\right)^{\nu_{12345}-D}
 \int \frac{d^Dk_1}{i \pi^{\frac{D}{2}}} \frac{d^Dk_2}{i \pi^{\frac{D}{2}}}
 \frac{1}{D_1^{\nu_1} D_2^{\nu_2} D_3^{\nu_3} D_4^{\nu_4} D_5^{\nu_5}},
 \;\;\;
\eq
with the propagators
\bq
 D_1=k_1^2-m^2, \hspace{0.3cm}  
 D_2=k_2^2, \hspace{0.3cm}  
 D_3 = (k_1-k_2)^2-m^2, \hspace{0.3cm} 
 D_4=(k_1-p)^2, \hspace{0.3cm}  
 D_5 = (k_2-p)^2-m^2
\eq
and $\nu_{12345}=\nu_1+\nu_2+\nu_3+\nu_4+\nu_5$.
The internal momenta are denoted by $k_1$ and $k_2$, 
the internal mass by $m$ and the external momentum by $p$. 
In the following we set $\mu=m$ and 
\bq
 x & = & \frac{p^2}{m^2}.
\eq
The Feynman integrals of the kite family are then only functions of $D$ and $x$ and we simply write
\bq
 I_{\nu_1 \nu_2 \nu_3 \nu_4 \nu_5}\left( D, x \right)
 & = &
 I_{\nu_1 \nu_2 \nu_3 \nu_4 \nu_5}\left( D, x m^2, m^2, m^2 \right).
\eq
This family of Feynman integrals is interesting, since it has an elliptic sub-sector corresponding to the
sunrise topology
\bq
 I_{\nu_1 0 \nu_3 0 \nu_5}\left( D, x \right).
\eq
For the sunrise topology the relevant elliptic curve may be either identified from the Feynman parameter
representation or the maximal cut.
Note that the $j$-invariants of the two elliptic curves differ,
therefore they are not related by a modular $\mathrm{PSL}(2,{\mathbb Z})$-transformation.
The lattice generated by the periods of the elliptic curve obtained from the Feynman parameter representation
is a sub-lattice of the one generated by the periods of the elliptic curve obtained 
from the maximal cut.
We present here the extraction of the elliptic curve from the maximal cut \cite{Adams:2017ejb}.
The method based on the maximal cut generalises easily to more complicated elliptic topologies.
In particular, the maximal cut of a Feynman integral is a solution of the homogeneous differential equation
for this Feynman integral \cite{Primo:2016ebd}.
We consider the maximal cut of the sunrise integral in two space-time dimensions.
One finds \cite{Baikov:1996iu,Lee:2009dh,Kosower:2011ty,CaronHuot:2012ab,Frellesvig:2017aai,Bosma:2017ens,Harley:2017qut}
\bq
\label{maxcut}
 \mathrm{MaxCut}_{\mathcal C} \; I_{10101}\left(2,x\right)
 & = &
 \frac{u}{\pi^2}
 \int\limits_{\mathcal C} 
 \frac{dz}{z^{\frac{1}{2}} \left(z + 4 \right)^{\frac{1}{2}} \left[z^2 + 2 \left(1+x\right) z + \left(1-x\right)^2 \right]^{\frac{1}{2}}},
\eq
where $u$ is an (irrelevant) phase and ${\mathcal C}$ an integration contour.
The denominator of the integrand defines an elliptic curve, which we denote by $E$:
\bq
\label{E_73_maxcut}
 E
 & : &
 w^2 - z
       \left(z + 4 \right) 
       \left[z^2 + 2 \left(1+x\right) z + \left(1-x\right)^2 \right]
 \; = \; 0.
\eq
We denote the roots of the quartic polynomial in eq.~(\ref{E_73_maxcut}) by
\bq
 z_1 \; = \; -4,
 \;\;\;
 z_2 \; = \; -\left(1+\sqrt{x}\right)^2,
 \;\;\;
 z_3 \; = \; -\left(1-\sqrt{x}\right)^2,
 \;\;\;
 z_4 \; = \; 0.
\eq
We consider a neighbourhood of $x=0$ without the branch cut of $\sqrt{x}$ along the negative real axis.
The correct physical value is specified by Feynman's $i\delta$-prescription: $x\rightarrow x+i\delta$.
We further set
\bq
 k^2 
 \; = \;
 \frac{\left(z_3-z_2\right)\left(z_4-z_1\right)}{\left(z_3-z_1\right)\left(z_4-z_2\right)},
 & &
 k'{}^2
 \; = \;
 \frac{\left(z_2-z_1\right)\left(z_4-z_3\right)}{\left(z_3-z_1\right)\left(z_4-z_2\right)}.
\eq
A standard choice of periods is then
\bq
 \psi_{1}
 \; = \;
 \frac{4 K\left( k \right)}{\left(1+\sqrt{x}\right)^{\frac{3}{2}} \left(3-\sqrt{x}\right)^{\frac{1}{2}}},
 & &
 \psi_{2}
 \; = \;
 \frac{4i K\left( k' \right)}{\left(1+\sqrt{x}\right)^{\frac{3}{2}} \left(3-\sqrt{x}\right)^{\frac{1}{2}}}.
\eq
We denote the ratio of the two periods and the nome squared by
\bq
\label{def_tau}
 \tau
 \;\; = \;\;
 \frac{\psi_{2}}{\psi_{1}},
 & &
 q \;\; = \;\; e^{2 i \pi \tau}.
\eq
Eq.~(\ref{def_tau}) defines $\tau$ as a function of $x$.
It is advantageous to change the variable in the base manifold from $x$ to $\tau$.
In a neighbourhood of $x=0$ we may invert eq.~(\ref{def_tau}). This gives
\bq
\label{hauptmodul}
 x
 & = &
 9
 \frac{\eta\left(6\tau\right)^8 \eta\left(\tau\right)^4}
      {\eta\left(2\tau\right)^8 \eta\left(3\tau\right)^4},
\eq
where $\eta$ denotes Dedekind's eta-function.
For the Jacobian we have
\bq
 \frac{d\tau}{dx}
 & = &
 \frac{W}{\psi_1^2},
\eq
where the Wronskian $W$ is given by
\bq
\label{Wronskian_relation}
 W
 & = & 
 \psi_{1} \frac{d}{dx} \psi_{2} - \psi_{2} \frac{d}{dx} \psi_{1}
 \;\; = \;\;
 - \frac{6 \pi i}{x\left(x-1\right)\left(x-9\right)}.
\eq
The advantage of the change of variables from $x$ to $\tau$ is given by the fact, that in the variable $\tau$ all
integration kernels are modular forms.
More specifically, the integration kernels are modular forms of the congruence subgroup $\Gamma_1(6)$.
We introduce a basis $\{e_1,e_2\}$ for the modular forms of modular weight $1$ 
for the Eisenstein subspace ${\mathcal E}_1(\Gamma_1(6))$:
\bq
 e_1 \; = \; E_1\left(\tau;\chi_0,\chi_1\right),
 & &
 e_2 \; = \; E_1\left(2\tau;\chi_0,\chi_1\right),
\eq
where $\chi_0$ and $\chi_1$ denote primitive Dirichlet characters with conductors $1$ and $3$, respectively.
The Eisenstein series $E_1(\tau,\chi_0,\chi_1)$ and $E_1(2\tau,\chi_0,\chi_1)$ 
are defined in appendix~\ref{sec:eisenstein}.
All occurring integration kernels may be expressed as polynomials in $e_1$ and $e_2$. 
We set
\bq
\label{integration_kernels}
% f_{1}
% & = &
% \frac{\left(x+3\right)}{2 \sqrt{6}} \; \frac{\psi_{1}}{\pi}
% \; = \;
% 3 \sqrt{2} e_1,
% \nonumber \\
 f_{2} 
 & = &
 \frac{1}{2 i \pi} \frac{\psi_{1}^2}{W} \frac{\left(3x^2-10 x - 9 \right)}{2 x \left(x-1\right) \left(x-9\right)}
 \; = \;
 -6 \left( e_1^2 + 6 e_1 e_2 - 4 e_2^2 \right),
 \nonumber \\
 f_{3} 
 & = &
 \frac{\psi_{1}^3}{4 \pi W^2}
 \;
 \frac{6}{x \left(x-1\right)\left(x-9\right)}
 \; = \;
 36 \sqrt{3} \left( e_1^3 - e_1^2 e_2 - 4 e_1 e_2^2 + 4e_2^3 \right),
 \nonumber \\
 f_{4}
 & = &
 \frac{1}{576}
 \frac{\psi_{1}^4}{\pi^4} \; \left(x+3\right)^4 
 \; = \;
 324 e_1^4,
 \nonumber \\
 g_{2,0} 
 & = &
 \frac{1}{2 i \pi} \frac{\psi_{1}^2}{W} \frac{1}{x}
 \; =  \;
 - 12 \left( e_1^2 - 4 e_2^2 \right),
 \nonumber \\
 g_{2,1} 
 & = &
 \frac{1}{2 i \pi} \frac{\psi_{1}^2}{W} \frac{1}{x-1}
 \; = \;
 - 18 \left( e_1^2 + e_1 e_2 - 2 e_2^2 \right),
 \nonumber \\
% g_{2,9} 
% & = &
% \frac{1}{2 i \pi} \frac{\psi_{1}^2}{W} \frac{1}{x-9}
% \; = \;
% 6 \left( e_1^2 - 3 e_1 e_2 + 2 e_2^2 \right),
% \nonumber \\
 g_{3,0} 
 & = &
 \frac{1}{2 i \pi} \frac{\psi_{1}^2}{W} \frac{\psi_{1}}{\pi}
 \; = \;
 -72 \sqrt{3} \left( e_1^3 - e_1^2 e_2 - 4 e_1 e_2^2 + 4e_2^3 \right),
 \nonumber \\
 g_{3,1} 
 & = &
 \frac{1}{2 i \pi} \frac{\psi_{1}^2}{W} \frac{\psi_{1}}{\pi} \frac{x}{x-1}
 \; = \;
 - 108 \sqrt{3} \left( e_1^3 - 3 e_1 e_2^2 + 2 e_2^3 \right).
\eq
The integration kernels $f_k$ and $g_{k,j}$ are modular forms of modular weight $k$ for $\Gamma_1(6)$.
Note that the integration kernels are not linear independent. We have for example
$g_{3,0}=-2 f_3$.
These integration kernels are supplemented by the constant function $1$, which is trivially
a modular form of modular weight $0$.
Let us further note that $\psi_1/\pi$ is a modular form of weight $1$:
\bq
 \frac{\psi_1}{\pi} 
 & = &
 2 \sqrt{3}
 \left( e_1 + e_2 \right).
\eq
If $f_1(\tau)$, $f_2(\tau)$, ..., $f_n(\tau)$ are modular forms and $f_n(\tau)$ vanishes at the cusp $\tau=i\infty$ we define
the $n$-fold iterated integral by
\bq
 I\left(f_1,f_2,...,f_n;q\right)
 & =
 \left(2 \pi i \right)^n
 \int\limits_{i \infty}^{\tau} d\tau_1
 \;
 f_1\left(\tau_1\right)
 \int\limits_{i \infty}^{\tau_1} d\tau_2
 \;
 f_2\left(\tau_2\right)
 ...
 \int\limits_{i \infty}^{\tau_{n-1}} d\tau_n
 \;
 f_n\left(\tau_n\right),
 \;\;\;\;\;\;
 q \; = \; e^{2\pi i \tau}.
\eq
The case where $f_n(\tau)$ does not vanishes at the cusp $\tau=i\infty$ is discussed in \cite{Adams:2017ejb,Brown:2014aa}
and is similar to trailing zeros in the case of multiple/harmonic polylogarithms.

For the integral family of the kite integral there are eight master integrals.
We may choose them as $\vec{I}=(I_1,I_2,I_3,I_4,I_5,I_6,I_7,I_8)^T$ with
\bq
\label{def_basis}
 I_1
 & = &
 4 \eps^2 I_{20200}\left(4-2\eps,x\right),
 \nonumber \\
 I_2
 & = &
 4 \eps^2 x I_{20210}\left(4-2\eps,x\right),
 \nonumber \\
 I_3
 & = &
 4 \eps^2 x I_{02210}\left(4-2\eps,x\right),
 \nonumber \\
 I_4
 & = &
 4 \eps^2 \left[ 2 I_{02210}\left(4-2\eps,x\right) + \left(1-x\right) I_{02120}\left(4-2\eps,x\right) \right],
 \nonumber \\
 I_5
 & = &
 4 \eps^2 x^2 I_{21012}\left(4-2\eps,x\right),
 \nonumber \\
 I_6
 & = &
 \eps^2 \frac{\pi}{\psi_1} I_{10101}\left(2-2\eps,x\right),
 \nonumber \\
 I_7
 & = &
 \eps \frac{i}{2 \psi_1^2} \left( \frac{d\psi_1 }{d\tau} \right) I_{10101}\left(2-2\eps,x\right)
 + \eps \frac{i \psi_1}{2 W} \left[ \frac{1}{x} I_{10101}\left(2-2\eps,x\right) - \frac{3}{x} I_{20101}\left(2-2\eps,x\right) \right]
 \nonumber \\
 & &
 - \eps^2 \frac{i \psi_1}{2 W} \left[ \frac{1}{x-1} + \frac{1}{x-9} - \frac{5}{2x} \right] I_{10101}\left(2-2\eps,x\right),
 \nonumber \\
 I_8 
 & = & - 8 \eps^3 \left(1-2\eps\right) x I_{11111}\left(4-2\eps,x\right).
\eq
A few comments are in order:
The first five master integrals correspond to the choice made in \cite{Remiddi:2016gno,Adams:2016xah}.
These master integrals can be expressed as harmonic polylogarithms in the variable $x$.
As they appear as sub-topologies for the kite integral, we treat them on the same footing as the remaining integrals.
The system of differential equations for these integrals with respect to the variable $\tau$ gives 
-- as for the remaining integrals -- integration kernels which are modular forms.

The master integral $I_6$ is basically the sunrise integral in $D=2-2\eps$ dimensions divided 
by the $\eps^0$-term of the corresponding maximal cut.
We recall that the period $\psi_1$ is (up to a trivial prefactor) equal to the maximal cut for a specific integration
contour for the cut integral.
The definition of $I_6$ is not unexpected. 
Experience from Feynman integrals evaluating to multiple polylogarithms supports the conjecture that
Feynman integrals with constant leading singularities will evaluate to iterated integrals of uniform length in each order of $\eps$.
Note that changing the master integral from $I_{10101}(2-2\eps,x)$ to $I_6$ is not a rational or algebraic transformation.
The period $\psi_1$ is a transcendental function.
It is exactly this extension of basis transformations in the fibre from algebraic/rational transformations to
transcendental transformations which allows us to obtain an $\eps$-form.
The required transformation is quite natural: We divide by the $\eps^0$-term of the maximal cut.
It is a simple fact that for the sunrise integral the $\eps^0$-term of the maximal cut is not an algebraic function, 
but a transcendental function.

A new result of this letter is an appropriate definition of the second master integral in the sunrise sector.
Our goal is to avoid the appearance of quasi-modular forms \cite{Kaneko:1995aa,Matthes:2017aa}.
To motivate the definition of $I_7$ let us first introduce
\bq
 L_{\mathrm{inhom}} & = & 
 \ln\left( \frac{\left(x-1\right)\left(x-9\right)}{3 \sqrt{-x}} \right)
 \; = \;
 I\left(f_2;q\right).
\eq
The result for the Feynman integral $I_{10101}(2-2\eps,x)$ has the form \cite{Adams:2017ejb}
\bq
 I_{10101}\left(2-2\eps,x\right)
 & = &
 \frac{\psi_1}{\pi}
 e^{-\eps L_{\mathrm{inhom}}}
 \Gamma\left(1+\eps\right)^2
 \tilde{E}_{111}\left(2-2\eps,q\right),
\eq
where the Taylor expansion $\tilde{E}_{111}$ has at each order $\eps^l$ iterated integrals of uniform length $(l+2)$. 
The second master integral $I_7$ for the sunrise sector is basically the $\tau$-derivative of $\tilde{E}_{111}$.
More concretely, we have
\bq
\label{I7_tau_derivative}
 I_7
 & = &
 \eps e^{-\eps L_{\mathrm{inhom}}} 
 \Gamma\left(1+\eps\right)^2 
 \frac{1}{2\pi i} 
 \frac{d}{d\tau} \tilde{E}_{111}\left(2-2\eps,q\right) 
 \nonumber \\
 & = & 
 \eps e^{-\eps L_{\mathrm{inhom}}} 
 \frac{1}{2\pi i} 
 \frac{d}{d\tau} \left( e^{\eps L_{\mathrm{inhom}}} \frac{\pi}{\psi_1} I_{10101}\left(2-2\eps,x\right) \right).
\eq
The equivalence of eq.~(\ref{I7_tau_derivative}) with the expression appearing in eq.~(\ref{def_basis}) 
is shown in appendix~\ref{sect:second_master_integral}.
In the definition of $I_6$ and $I_7$ we used Feynman integrals in $D=2-2\eps$ dimensions.
Using dimensional shift relations \cite{Tarasov:1996br,Tarasov:1997kx}, 
the integrals are easily expressed in terms of Feynman integrals in $D=4-2\eps$ dimensions.
The relevant formulae are given in appendix~\ref{sec:dim_shift}.

The choice of the master integral $I_8$ corresponds to the choice made in \cite{Remiddi:2016gno}.

In the basis of eq.~(\ref{def_basis}) the system of differential equations for the kite family is in $\eps$-form.
We have
\bq
\label{res_eps_form}
 \frac{1}{2\pi i} \frac{d}{d\tau} \vec{I}
 & = &
 \eps \; A \; \vec{I},
\eq
where the matrix $A$ is $\eps$-independent and contains only the integration kernels of eq.~(\ref{integration_kernels})
and the trivial constant modular form $1$.
Explicitly we have
\bq
\label{res_A}
 A & = & 
 \left( \begin{array}{rrrrrrrr}
 0 & 0 & 0 & 0 & 0 & 0 & 0 & 0 \\
 -g_{2,1} & g_2 & 0 & 0 & 0 & 0 & 0 & 0 \\
 0 & 0 & g_2 & g_{2,1} & 0 & 0 & 0 & 0 \\
 0 & 0 & -4 g_{2,0} + 4 g_{2,1}& -2 g_{2,1} & 0 & 0 & 0 & 0 \\
 0 & -2 g_{2,1} & 0 & 0 & 2 g_2 & 0 & 0 & 0 \\
 0 & 0 & 0 & 0 & 0 & -f_2 & 1 & 0 \\
 \frac{1}{4} f_3 & 0 & 0 & 0 & 0 & f_4 & -f_2 & 0 \\
 g_{2,1} & 0 & -2 g_{2,1} & -g_{2,1} & -2 g_{2,0} & -12 g_{3,0} + \frac{32}{3} g_{3,1}  & 0 & g_2 \\
 \end{array} \right).
 \nonumber \\
\eq
As abbreviation we used
\bq
 g_2 & = & g_{2,0} - 2 g_{2,1}.
\eq
Eq.~(\ref{res_eps_form}) and eq.~(\ref{res_A}) are the main results of this letter
and give the $\eps$-form of the system of differential equations for the kite family.
This system of differential equations is easily integrated order-by-order in $\eps$.
The integration kernels appearing in eq.~(\ref{res_A}) are all modular forms of the congruence subgroup
$\Gamma_1(6)$.
The $\eps$-form is obtained by a (transcendental) change of basis of master integrals as given in eq.~(\ref{def_basis}).
Let us mention that the (transcendental) change of variables from $x$ to $\tau$ is not required to
obtain an $\eps$-form.
This is easily seen by first noticing that the change of variables from $x$ to $\tau$ is $\eps$-independent
and secondly transforming eq.~(\ref{res_eps_form}) and eq.~(\ref{res_A}) back to $x$.
However, let us stress that we are not only interested in obtaining an $\eps$-form for the system of differential
equations, at the same time we would also like to achieve that the integration kernels belong to a nice class of
functions.
Not performing the change of variables from $x$ to $\tau$ would hide the fact that the integration kernels 
are for a suitable variable modular forms of $\Gamma_1(6)$.

% -----------------------------------------------------------------------------
\section{Conclusions}
\label{sec:conclusions}

In this letter we have shown that the system of differential equations for certain Feynman integrals may
be transformed to an $\eps$-form, even if these Feynman integrals cannot be expressed in terms of multiple polylogarithms.
This can be achieved by allowing transcendental functions as entries of the transformation matrix $U$, which defines the 
basis change in the fibre.

For the concrete example of the kite integral, a change of variables in the base manifold
from $x$ to $\tau$ turns all integration kernels into
modular forms of the congruence subgroup $\Gamma_1(6)$.

With known boundary conditions the resulting system of differential equations is then easily integrated to any desired order in $\eps$.

We expect our results to be useful for the further development of the theory of elliptic generalisations of
multiple polylogarithms \cite{Beilinson:1994,Levin:1997,Levin:2007,Enriquez:2010,Brown:2011,Wildeshaus,Bloch:2013tra,Bloch:2014qca,Adams:2014vja,Adams:2015gva,Adams:2015ydq,Adams:2016xah,Passarino:2017EPJC,Remiddi:2017har,Broedel:2017kkb,Broedel:2017siw},
and for more complicated Feynman integrals appearing 
in precision calculations in high-energy physics \cite{Adams:2014vja,Adams:2015gva,Bonciani:2016qxi,vonManteuffel:2017hms,Primo:2017ipr,Ablinger:2017bjx,Bourjaily:2017bsb,Hidding:2017jkk}
and string theory \cite{Broedel:2014vla,Broedel:2015hia,Broedel:2017jdo,D'Hoker:2015qmf,Hohenegger:2017kqy}.

\subsection*{Acknowledgements}

We would like to thank the anonymous referee of \cite{Adams:2017ejb} for useful hints and suggestions.

% -----------------------------------------------------------------------------
\begin{appendix}

\section{Eisenstein series}
\label{sec:eisenstein}

In this appendix we give the explicit expressions for the 
Eisenstein series $E_1(\tau,\chi_0,\chi_1)$ and $E_1(2\tau,\chi_0,\chi_1)$.
$\chi_0$ and $\chi_1$ denote primitive Dirichlet characters with conductors $1$ and $3$, respectively.
In terms of Kronecker symbols they are given by
\bq
 \chi_0 \; = \; \left( \frac{1}{n} \right), & &
 \chi_1 \; = \; \left( \frac{-3}{n} \right).
\eq
More explicitly we have
\bq
 \chi_0\left(n\right) & = & 1, \;\;\;\;\;\; \forall n \in {\mathbb Z},
 \nonumber \\
 \chi_1\left(n\right) 
 & = & 
 \left\{ \begin{array}{rl}
  0, & n = 0 \mod 3, \\
  1, & n = 1 \mod 3, \\
 -1, & n = 2 \mod 3, \\
 \end{array} \right.
\eq
$E_1(\tau,\chi_0,\chi_1)$ is given with $q=e^{2\pi i \tau}$ by
\bq
 E_1\left(\tau; \chi_0, \chi_1\right) 
 & = &
 \frac{1}{6} + \sum\limits_{m=1}^{\infty} \left( \sum\limits_{d|m} \chi_1\left(d\right) \right) q^m.
\eq
In terms of the $\mathrm{ELi}$-functions, defined by
\bq
 \mathrm{ELi}_{n;m}\left(x;y;q\right) 
 & = &
 \sum\limits_{j=1}^\infty \sum\limits_{k=1}^\infty \; \frac{x^j}{j^n} \frac{y^k}{k^m} q^{j k},
\eq
we have
\bq
 E_1\left(\tau; \chi_0, \chi_1\right) 
 & = &
 \frac{1}{6} 
 + \frac{1}{i \sqrt{3}} \left[ \mathrm{ELi}_{0,0}\left(r_3,1;q\right) - \mathrm{ELi}_{0,0}\left(r_3^{-1},1;q\right) \right],
\eq
where $r_3=\exp(2\pi i / 3)$ denotes the third root of unity.
The first few terms of $E_1(\tau,\chi_0,\chi_1)$ read
\bq
 E_1\left(\tau; \chi_0, \chi_1\right) 
 & = &
 \frac{1}{6} + q + q^3 + q^4 + 2 q^7 + q^9 + ...
\eq
The Eisenstein series $E_1(2\tau,\chi_0,\chi_1)$ is obtained from $E_1(\tau,\chi_0,\chi_1)$ by
the substitution $\tau \rightarrow 2 \tau$ or equivalently $q \rightarrow q^2$.

\section{Dimensional shift relations}
\label{sec:dim_shift}

The integrals $I_{10101}(2-2\eps,x)$ and $I_{20101}(2-2\eps,x)$ in $D=2-2\eps$ space-time dimensions are easily
expressed in terms of Feynman integrals in $D=4-2\eps$ space-time dimensions.
We have
\bq
 I_{10101}\left(2-2\eps,x\right)
 & = &
 3 I_{20201}\left(4-2\eps,x\right),
 \nonumber \\
 I_{20101}\left(2-2\eps,x\right)
 & = &
 4 I_{30201}\left(4-2\eps,x\right)
 + I_{20202}\left(4-2\eps,x\right).
\eq
Reducing them to a basis of master integrals in $D=4-2\eps$ space-time dimensions gives
\bq
\lefteqn{
 I_{10101}\left(2-2\eps,x\right)
 = 
 \frac{3 \left(3-x\right)}{\left(x-1\right)\left(x-9\right)} I_{20200}\left(4-2\eps,x\right)
 + \frac{6 \left(1-2\eps\right)\left(2-3\eps\right)}{\left(x-1\right)\left(x-9\right)} I_{10101}\left(4-2\eps,x\right)
 } & &
 \nonumber \\
 & &
 + \frac{6 \left(1-2\eps\right)\left(x+3\right)}{\left(x-1\right)\left(x-9\right)} I_{20101}\left(4-2\eps,x\right),
 \nonumber \\
\lefteqn{
 I_{20101}\left(2-2\eps,x\right)
 =
 \left[ \frac{3}{\left(x-1\right)\left(x-9\right)} - 2 \eps \frac{x^3-17x^2+27x-27}{\left(x-1\right)^2\left(x-9\right)^2} \right]I_{20200}\left(4-2\eps,x\right)
 } & &
 \nonumber \\
 & &
 + 2 \left(1-2\eps\right)\left(2-3\eps\right) \left[ \frac{1}{\left(x-1\right)\left(x-9\right)} - 2 \eps \frac{x^2-9}{\left(x-1\right)^2\left(x-9\right)^2}\right] I_{10101}\left(4-2\eps,x\right)
 \nonumber \\
 & &
 + 2 \left(1-2\eps\right) \left[ \frac{3}{\left(x-1\right)\left(x-9\right)} + \eps \frac{x^3-36x^2+45x+54}{\left(x-1\right)^2\left(x-9\right)^2}\right] I_{20101}\left(4-2\eps,x\right).
\eq

\section{The second master integral in the sunrise sector}
\label{sect:second_master_integral}

In this appendix we show the equivalence of the definition of $I_7$ 
given in eq.~(\ref{I7_tau_derivative})
\bq
\label{I7_def1}
 I_7
 & = &
 \eps e^{-\eps L_{\mathrm{inhom}}} 
 \frac{1}{2\pi i} 
 \frac{d}{d\tau} \left( e^{\eps L_{\mathrm{inhom}}} \frac{\pi}{\psi_1} I_{10101}\left(2-2\eps,x\right) \right)
\eq
with the one appearing eq.~(\ref{def_basis})
\bq
\label{I7_def2}
 I_7
 =
 \eps \frac{i}{2 \psi_1^2} \left( \frac{d\psi_1 }{d\tau} \right) I_{10101}\left(2-2\eps,x\right)
 + \eps \frac{i \psi_1}{2 W} \left[ \frac{1}{x} I_{10101}\left(2-2\eps,x\right) - \frac{3}{x} I_{20101}\left(2-2\eps,x\right) \right].
\eq
Let us start from eq.~(\ref{I7_def1}).
Applying the product rule for differentiaition we obtain
\bq
 I_7
 & = &
 \eps^2 
 \frac{\pi}{\psi_1} I_{10101}\left(2-2\eps,x\right) 
 \frac{1}{2\pi i} 
 \frac{d}{d\tau} L_{\mathrm{inhom}}
 -
 \eps 
 \frac{\pi^2}{\psi_1^2}
 I_{10101}\left(2-2\eps,x\right)   
 \frac{1}{2\pi i} 
 \frac{d}{d\tau} \left( \frac{\psi_1}{\pi} \right)
 \nonumber \\
 & &
 +
 \eps \frac{\pi}{\psi_1} 
 \frac{1}{2\pi i} 
 \frac{d}{d\tau} I_{10101}\left(2-2\eps,x\right).
\eq
In the first term we have
\bq
 \frac{1}{2\pi i} 
 \frac{d}{d\tau} L_{\mathrm{inhom}}
 \; = \;
 \frac{1}{2\pi i} 
 \frac{d}{d\tau} I\left(f_2;q\right)
 \; = \; 
 f_2
 \; = \;
 \frac{1}{2 i \pi} \frac{\psi_{1}^2}{W} \frac{\left(3x^2-10 x - 9 \right)}{2 x \left(x-1\right) \left(x-9\right)}.
\eq
In the third term we replace the differentiation with respect to $\tau$ by the differentiation with respect to $x$
and use the differential equation for $I_{10101}$:
\bq
 \frac{1}{2\pi i} 
 \frac{d}{d\tau} I_{10101}\left(2-2\eps,x\right)
 & = &
 \frac{\psi_1^2}{2\pi i W} 
 \frac{d}{dx} I_{10101}\left(2-2\eps,x\right)
 \nonumber \\
 & = &
 \frac{\psi_1^2}{2\pi i W} 
 \left[
  \frac{3}{x} I_{20101}\left(2-2\eps,x\right)
  -
  \frac{1+2\eps}{x} I_{10101}\left(2-2\eps,x\right)
 \right].
\eq
Putting everyhing together gives eq.~(\ref{I7_def2}).

\end{appendix}

%------------------------------------------------------------------------------
% references
{\footnotesize
\bibliography{/home/stefanw/notes/biblio}
\bibliographystyle{/home/stefanw/latex-style/h-physrev5}
}

\end{document}